\documentclass[amssymb,amsmath,aps,secnumarabic,floatfix,twocolumn,pre,showpacs]{revtex4}
\usepackage{epsfig}
\usepackage{xspace}

\newcommand{\latin}[1]{{\it #1}}
\newcommand{\ie}{\latin{i.e.}\@\xspace}

\makeatletter
\newcommand{\supmarker}[1]{{\@ifempty{#1}{}{\text{(#1)}}}}
\makeatother
\newcommand{\pdf}[2]{\mathcal{P}^{\supmarker{#1}}\left(#2\right)}

\newcommand{\elabel}[1]{\label{eq:#1}}

\newcommand{\Eref}[1]{Eq.~(\ref{eq:#1})}

\newcommand{\flabel}[1]{\label{fig:#1}}

\newcommand{\Fref}[1]{Fig.~\ref{fig:#1}}

\newcommand{\tlabel}[1]{\label{table:#1}}

\newcommand{\Tref}[1]{Table~\ref{table:#1}}

\newcommand{\GC}{\mathcal{G}}

\newcommand{\plaind}{\mathrm{d}}
\newcommand{\dint}[1]{\mathchoice{\!\plaind#1\,}{\!\plaind#1\,}{\!\plaind#1\,}{\!\plaind#1\,}}

\newcommand{\ave}[1]{\left\langle #1 \right\rangle}

\expandafter\ifx\csname package@font\endcsname\relax\else
  \expandafter\expandafter
  \expandafter\usepackage
  \expandafter\expandafter
  \expandafter{\csname package@font\endcsname}%
\fi

\begin{document}
\title{The Abelian Manna model on two fractal lattices}
\date{\today}
\author{Hoai Nguyen Huynh}
\affiliation{Division of Physics and Applied Physics,
School of Physical and Mathematical Sciences,
Nanyang Technological University, Singapore,
21 Nanyang Link, Singapore 637371}

\author{Gunnar Pruessner}
\affiliation{Department of Mathematics,
Imperial College London,
180 Queen's Gate,
London SW7 2BZ, United Kingdom}

\author{Lock Yue Chew}
\affiliation{Division of Physics and Applied Physics,
School of Physical and Mathematical Sciences,
Nanyang Technological University, Singapore,
21 Nanyang Link, Singapore 637371}

\begin{abstract}
We analyze the avalanche size distribution of the Abelian Manna
model on two different fractal lattices with the same dimension
$d_g=\ln3/\ln2$, with the aim to probe for scaling behavior and
to study the systematic dependence of the critical exponents
on the dimension and structure of the lattices. 
We show that the scaling law $D(2-\tau)=d_w$ generalizes the
corresponding scaling law on regular lattices, in particular hypercubes,
where $d_w=2$.
Furthermore, we observe that the lattice dimension
$d_g$, the fractal dimension of the random walk on the
lattice $d_w$, and the critical exponent $D$, form a plane
in $3D$ parameter space, i.e. they obey the linear
relationship
$D=0.632(3)d_g+0.98(1)d_w-0.49(3)$.
\end{abstract}


\maketitle

Although extensive research has been performed on self-organized
criticality \cite{BakETAL:1987} for models on hypercubic lattices, 
far less work has been done on fractal lattices \cite{Kutnjak-UrbancETAL:1996,LeeETAL:2009}. 
It remains somewhat unclear what to conclude from the latter studies.
Fractal lattices are important for the understanding of critical phenomena
for a number of reasons. Firstly, results for critical exponents
in lattices with non-integer dimensions might provide a means to
determine the terms of their $\epsilon=4-d$ expansion. Secondly,
fractal lattices are particularly suitable for a real space
renormalization group procedures, in particular that by Migdal and
Kadanoff \cite{Migdal:1976,Kadanoff:1976,CarmonaETAL:1998}. Thirdly,
scaling relations that are derived in a straightforward fashion on
hypercubic lattices can be put to test in a more general setting.
In this Brief Report, we address the first and the third aspect, by
examining both numerically and analytically
the scaling behaviour of the Abelian version of the Manna
model \citep{Manna:1991a,
Dhar:1999,
Dhar:2006} on two different
fractal lattices.

The fractal lattices used in this study are generated from the
arc-fractal system \cite{HuynhChew:2010}. The lattice sites are
the invariant set of points of the arc-fractal. We consider nearest
neighbor interactions among sites. 
Here, the nearest neighbors of a given site are all sites which
have the (same) shortest Euclidean
distance to it.
Our fractal lattices have no
natural boundary; instead, they have only two end points at
which two copies can join to form a bigger lattice. The dimension
of the lattices is the same as the arc-fractal that generates
them.

In this study, we shall consider two fractal lattices: the Sierpinski
arrowhead and the crab (see \Fref{lattices}). The former is named ``Sierpinski arrowhead''
because it is the same as the well-known Sierpinski
arrowhead \cite{Mandelbrot:1982}, whereas the latter is termed ``crab''
because the overall shape of the generated lattice looks like a crab.
These fractal lattices are generated through the arc-fractal system
with number of segments $n=3$ and opening angle of the arc $\alpha=\pi$.
For the Sierpinski arrowhead, the rule for orientating the arc at each
iteration is ``in-out-in'', while the rule is ``out-in-out'' for the
crab. 
Both 
lattices have the same dimension
$d_g=\ln3/\ln2\approx1.58$. The total number of sites on the
lattice at the $i$-th iteration is $N_i=3^i+1$. The coordination number
of sites on these lattices varies between two and three. Asymptotically,
one-third of the sites have three nearest neighbors
(called extended sites), while the remaining 
two thirds 
have two
nearest neighbors (called normal sites). Since the lattice sites are
being stringed up by arcs (see \Fref{lattices}), they can be labeled as sites
on a $1D$ linear chain. In order to determine the linear size 
(to be used in finite size scaling) 
of the
lattice, reference sites (hollow circles, which are not part of the
lattice) have been added between the real sites (solid circles). This
is possible because of the uniform spacing between sites. The linear
size is then equal to the total number of hollow and solid circles
along $L$ (see \Fref{lattices}). At the $i$-th iteration, the linear size is
given by $L_i=2^i+1$ for the arrowhead lattice and
$L_i=3\times2^{i-1}$ for the crab lattice. Indeed, one observes that
the dimension of the lattice obeys the following relation with the
total number of sites and linear size of the lattice:
$d_g=\lim_{i\rightarrow\infty}\ln{N_i}/\ln{L_i}$.

\begin{figure}[h]
  \centering
      \includegraphics[width=0.3\textwidth]{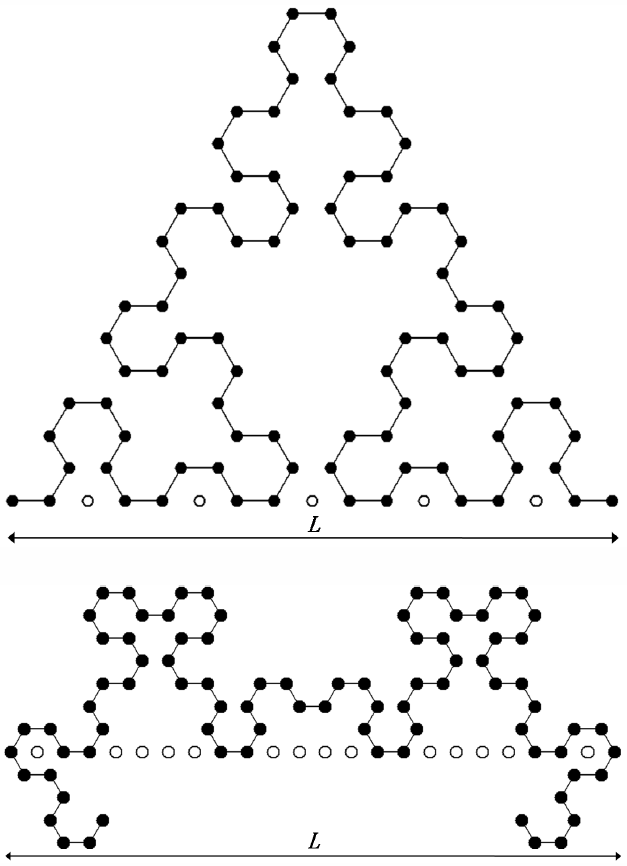}
  \caption{Sierpinski arrowhead (top) and crab (bottom) lattice at
  $4^{th}$ iteration with definition of linear size. In the figure,
  $L=17$ for arrowhead and $L=24$ for crab lattice.
  \flabel{lattices}}
\end{figure}

We implement the Abelian Manna model on these two fractal
lattices that have the same dimension but possess different
microscopic structures. Let us denote the variable $z_j$, which
is a non-negative integer, to be the ``height'' or the number of
particles at lattice site $j$. The lattice is first initialized
with $z_j=0$ for all sites. The value of $z_j$ is then evolved
according to the following algorithm. When the system is in a
stable configuration, i.e. $z_j \leq 1 $ for all sites, the external
drive is implemented by picking a site $j$ at random and increasing
its $z_j$ by one unit. 
At this juncture, an avalanche might occur
in the following manner: as long as there exist any $j$ with
$z_j$ exceeding the threshold $z_c=1$ (an ``active'' site), pick
one of them at random, say $k$, and reduce $z_k$ by two. At the
same time, pick two of its nearest neighbours, say $k'$, randomly
and independently, and increment $z_{k'}$ by one unit.
This procedure constitutes a toppling, which in the bulk is
conservative, \ie the total $\sum_j z_j$ remains unchanged
by the bulk topplings. In one dimension, every
bulk site has two neighbours. On the fractal lattices described
above, a bulk site can have either two or three nearest neighbours.
Note that 
particles
can leave the system at the two end sites (labeled as
$1$ and $3^i+1$ respectively). 
Owing to the Abelianess of the model, the statistics of the avalanche
sizes is independent from the order of updates. This is different for
time-dependent observables such as the avalanche duration, which is not
studied in the following.

An avalanche ceases as soon as $z_j \le 1$ for all $j$ (quiescence).
The size $s$ of the avalanche is measured as the number of
topplings performed between two quiescent configurations. The
probability density $\pdf{}{s}$ for an avalanche of size $s$ to
occur is expected to follow simple (finite size) scaling
\begin{equation}
\pdf{}{s} = a s^{-\tau} \GC\left(\frac{s}{b L^D}\right)
\elabel{simple_scaling_s}
\end{equation}
asymptotically in large $s\gg s_0$
with lower cutoff $s_0$, linear system size $L$,
non-universal metric factors $a$ and $b$ and universal
exponents $\tau$ and $D$. The universal scaling function $\GC$ decays,
for large arguments, faster than any power law, so that all moments
\begin{equation}
\ave{s^q} = \int \dint{s} \pdf{}{s} s^q
\elabel{def_moments}
\end{equation}
exist for any finite system. Provided that $q+1-\tau>0$ one can easily
show that
\begin{equation}
\ave{s^q} \propto L^{D(q+1-\tau)}  \ .
\end{equation}

The results presented in the following are based on Monte Carlo simulations for four different
system sizes, corresponding to four different levels of iteration
$i=3,4,5,6$, containing $N=28,82,244,730$ sites and with linear size
$L=9,17,33,65$ for the Sierpinski arrowhead lattice and
$L=12,24,48,96$ for the crab lattice. In all four cases, $10^8$
avalanches were triggered and the data recorded in regular intervals
of $10^6$ avalanches. Stationarity was verified by inspecting avalanche
size moments and the transient was determined to be shorter than
$5\cdot 10^4$ avalanches. Errors for the moments are derived from a
Jackknife estimator \cite{Efron:1982,Berg:1992} of the variance based on
the moments taken in each set of $10^6$ measurements.
The
scaling exponents were determined by a nonlinear least square
fitting \cite{PressETAL:2007} of the avalanche size moments against
the linear size of the lattice:
\begin{equation}
\ave{s^q} = \left(a_q + \frac{b_q}{L} + \frac{c_q}{L^2}\right) L^{D(q+1-\tau)}  \ .
\elabel{moment_scaling_fit}
\end{equation}
The exponents derived in this procedure were $\tau=1.170(5)$ and
$D=2.792(2)$ for the arrowhead lattice; and $\tau=1.153(4)$ and
$D=3.026(2)$ for the crab lattice.

In addition, the scaling behaviour can be probed in a data
collapse, as illustrated in \Fref{data_collapse}.

\begin{figure}[h]
  \centering
      \includegraphics[width=0.4\textwidth]{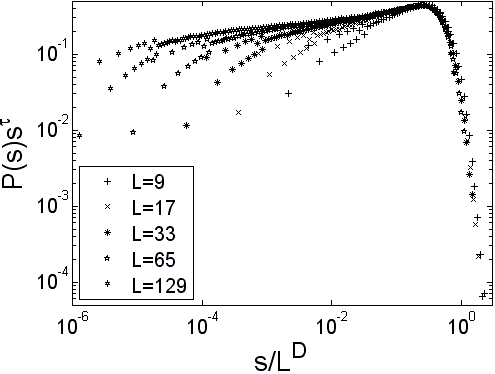}
  \caption{Data collapse of binned data for Sierpinski arrowhead 
  lattice. Preliminary statistics of large system size ($7^{th}$
  iteration, $L=129$) 
  are also included to test the consistency of the data. The
  plots confirm the estimated values of the critical exponents.
  \flabel{data_collapse}}
\end{figure}

With the critical exponents determined, one can immediately verify
the scaling law of avalanche size distribution. 
From hypercubic
lattices it is well known \cite{NakanishiSneppen:1997} that the
first moment of the avalanche size is given by the expected number of
moves that a random walker performs on the given lattice before it
reaches the boundary and leaves, \ie by its residence time.
This is essentially because of bulk conservation: In the stationary
state one particle leaves the system for every particle added (avalanche
attempt) and the average number of moves it performs during its
residency is exactly twice the average number of topplings occuring in
the system per particle added, which is the avalanche size.
Regardless of the specifics of the boundary, \ie regardless of
whether only two sites are dissipative or all sites along the
perimeter, the first moment normally scales with the linear
size of the lattices squared, $D(2-\tau)=2$, \emph{independent
from the dimension of the hypercubic lattice}. This is easily
understood, as the time and thus the total number of moves
performed by a random walker scales quadratically in the linear
distance traversed.

For the fractal lattices, it is obvious that $D(2-\tau)$ is not
equal to 2 since it is $2.317(8)$ for the arrowhead lattice and
$2.564(6)$ for the crab lattice. We will now show that the scaling
law has in fact changed to $D(2-\tau)=d_w$, where $d_w$ is the
fractal dimension of random walk on the lattice. 
The scale law
$D(2-\tau)=d_w$ remains true for any lattice regardless of
dimension and microscopic details.

We will now calculate $d_w$ for the Sierpinski arrowhead lattice and
the crab lattice by using the first-passage time
method \cite{AvrahamHavlin:2000}. 
Due to the nearest neighbor
structure of the fractal lattices, the calculation is
{\it not} coarse-grained renormalization-like, but rather,
it is carried out by considering every single edge that connects
between the two end sites on the lattice.

We label the sites of the lattice 
sequentially with
$j=1,2,\cdots,3^i+1$ and denote 
by $T_j^{(i)}$ the average time for
the random walker to exit through the end sites from site $j$ ($i$ is the
number of iterations of the lattice). 
Also, we denote by $t$ the
traverse time from one site to its nearest neighbor.

Let us define $E^{(i)}$ to be the set of extended sites at the $i$-th
iteration. On the lattice, each normal site $j$ has two nearest
neighbors $j-1$ and $j+1$. If $j$ is an extended site, it has in
addition a third nearest neighbor $j^*$. If
the walker is at site $j$, we have 2 situations:

If $j \not\in E^{(i)}$, the walker has only 2 options to choose from:
go to $j-1$ or go to $j+1$. The probability for each choice is $1/2$.
We have
\begin{eqnarray}
T_j^{(i)} &=& \frac{1}{2}\left(t+T_{j-1}^{(i)}\right)
+ \frac{1}{2}\left(t+T_{j+1}^{(i)}\right)
\nonumber\\
&=& t + \frac{1}{2}T_{j-1}^{(i)} + \frac{1}{2}T_{j+1}^{(i)}  \ .
\end{eqnarray}

If $j \in E^{(i)}$, the walker has up to 3 options to choose from: go
to $j-1$, go to $j+1$ or go to $j^*$. The probability for each choice
is $1/3$. Thus we have
\begin{eqnarray}
T_j^{(i)} &=& \frac{1}{3}\left(t+T_{j-1}^{(i)}\right) + \frac{1}{3}
\left(t+T_{j+1}^{(i)}\right) + \frac{1}{3}\left(t+T_{j^*}^{(i)}
\right) \nonumber\\
&=& t + \frac{1}{3}T_{j-1}^{(i)} + \frac{1}{3}T_{j+1}^{(i)}
+ \frac{1}{3}T_{j^*}^{(i)}  \ .
\end{eqnarray}

\noindent
Since this gives a system of linear equations, we can write them in
the form of a matrix equation: $A^{(i)}T^{(i)}=B^{(i)}$, where
$T^{(i)}$ is the column vector of $T_j^{(i)}$, 
$A^{(i)}$ is the degree-normalised adjacency matrix,
while $B^{(i)}$ is a column vector contains $t$ in every entry.
As $A^{(i)}$ is invertible,
we can solve for $T^{(i)}=\left(A^{(i)}\right)^{-1}B^{(i)}$.
By defining the new variable:
$S_j^{(i)}=T_j^{(i+1)}/T_j^{(i)}$, we obtain the
results as shown in 
\Tref{convergence_S}
for the arrowhead lattice and the crab
lattice.

\begin{table}
\caption{
\tlabel{convergence_S}
Convergence of $S_1^{(i)}$ as the number of iterations
$i$ increases. After $i=7$, we have $S_1^{(i)}\approx5$ for the
Sierpinski arrowhead lattice and $S_1^{(i)}\approx6$ for the crab
lattice. Other $S_j^{(i)}$'s are found to converge to the same limits.}
\begin{tabular}{cccccccc}
    \hline\hline
    $i$															&	1 	& 2 			& 3 			& 4 			& 5 		 	& 6				 & 7				\\
    \hline
    Arrowhead & 11	& 5.5076	& 4.7553	& 4.8806	& 4.9610	& 4.9893	&	4.9974	\\
    Crab & 10	& 7.1417	& 6.9528	&	6.3729	&	6.1014	&	6.0303	& 5.9727	\\
    \hline\hline
    \tlabel{convergence_S}
\end{tabular}
\end{table}

\begin{figure}[h]
  \centering
      \includegraphics[width=0.4\textwidth]{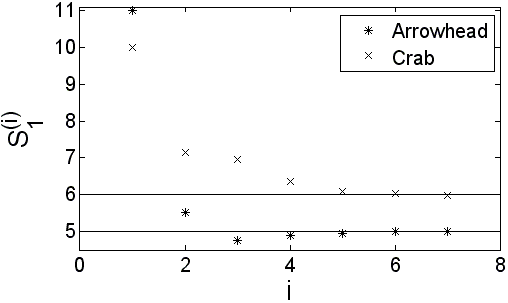}
  \caption{Convergence of $S_1^{(i)}$ as number of iterations  $i$ increases. After $i=7$ we have $S_1^{(i)}=5$ for arrowhead lattice and $S_1^{(i)}\approx6$ for crab lattice. Other $S_j^{(i)}$'s converge to the same limits.}
\end{figure}

We observe that for both lattices $S_j^{(i)}$ converges
to some value $S_j^*$ as $i$ increases. For the arrowhead
lattice, $S_j^*=5$, while for the crab lattice,
$S_j^*\approx6$. On going from the $i$-th to $(i+1)$-th
iteration, the lattice size 
increases by a factor 
$L_{i+1}/L_i=2$, while the escape time 
increases by a factor 
$T^{(i+1)}/T^{(i)}=S_j^*$ for large $i$,
so that
$d_w=\ln{S_j^*}/\ln2$. By comparing the calculated
$d_w$ to the estimated value of $D(2-\tau)$ from simulation, we
can see that they are in good agreement 
(see \Tref{dw_comparison}).

\begin{table}
\caption{
\tlabel{dw_comparison}
The calculated $d_w$ and the estimated $D(2-\tau)$ for the
Sierpinski arrowhead lattice and the crab lattice.}
\begin{tabular}{ccc}
    \hline\hline
    Lattice			& Arrowhead				& Crab						\\
    \hline
    $d_w$				& $2.322$ 				& $2.578$					\\
    $D(2-\tau)$	&	$2.317\pm0.008$	& $2.564\pm0.006$	\\
    \hline\hline
\end{tabular}
\end{table}

One surprising conclusion from the above results is that
for two lattices with the same dimension but different microscopic
structure, the critical exponents $D$ and $\tau$ 
can both be different, which
suggests that on fractal lattices, the critical exponents
depend not only on the dimension but also on the microscopic details
of the lattice. In addition, our results have validated the scaling
relation $D(2-\tau)=d_w$ for two fractal lattices with different
$d_w$. Since 
on lattices with integer dimensions 
$d_w=2$ regardless of the lattice's structure, this scaling
relation generalizes the standard version $D(2-\tau)=2$ known for
the hypercubic lattices. The last unexpected conclusion pertains
to the relation between the dimensions $d_g$ and $d_w$, and the
critical exponent $D$. We found that they obey the general
linear relationship: $D=\alpha d_g + \beta d_w + \gamma$. This was
uncovered by plotting the following six points in $3D$ parameter
space: $(d_g,d_w,D)=(1,2,2.1),(2,2,2.73),(3,2,3.36),(4,2,4)$,
$(1.58,2.32,2.79),(1.58,2.58,3.03)$, which correspond to
the linear chain \cite{NakanishiSneppen:1997}, the square
\cite{ChessaETAL:1999,
Manna:1991a},
the cube \cite{Pastor-SatorrasVespignani:2001,Ben-HurBiham:1996},
the hypercube \cite{Luebeck:2004}, the arrowhead and the crab
lattices respectively. It is interesting that while the first four
points 
due to the hypercubic lattices are found to lie
on a straight line, all 
six points together
make up a plane
instead of a tetrahedron. We have determined the coefficients
$\alpha=0.632\pm0.003$, $\beta=0.980\pm0.014$ and
$\gamma=-0.492\pm0.034$ from the six points, which leads
to the following relationship:
\begin{equation}
D \approx D_{trial}(d_g,d_w) = 0.632d_g+0.980d_w-0.492  \ .
\elabel{dg_dw_D}
\end{equation}
\noindent

\begin{table}
\caption{
\tlabel{D_table}
Data for $(d_g,d_w,D)$ from 6 different lattices and the calculated $D_{trial}$ from \Eref{dg_dw_D}.}
\begin{tabular}{ccccccc}
    \hline\hline
    Lattices		&	$1D$		&	$2D$		&	$3D$		&	$4D$		&	Arrowhead	&	Crab			\\
    \hline
    $d_g$ 			&	1 			& 2 			& 3 			& 4 			& 1.58 		 	& 1.58 			\\
    $d_w$ 			&	2 			& 2 			& 2 			& 2 			& 2.32 	 		& 2.58 			\\
    $D$ 				& 2.2(1) 	& 2.73(2)	& 3.36(1)	&	4				&	2.792(2)	&	3.026(2)	\\
    $D_{trial}$	&	2.10(7)	& 2.73(7)	& 3.36(7)	& 4.00(7)	&	2.78(7)		&	3.03(7)		\\
    \hline\hline
\end{tabular}
\end{table}

A comparison of the value of $D$ determined by \Eref{dg_dw_D}
and that from the simulation is shown in 
\Tref{D_table}.

Finally, 
we comment on
the slight mismatches between the calculated $d_w$ and
the estimated $D(2-\tau)$ in 
\Tref{dw_comparison}.
They seem to be caused by finite size corrections, which are further
suppressed at $7$ iterations and above, as observed in preliminary data
not included in the main analysis above. In the presence of strong
finite size corrections and high accuracy measurements of the moments,
the estimates for $D$ and $\tau$ are sensitive to the choice of the
fitting function, which is constrained by the number of
data points (\ie system sizes) available, but needs to contain as many
correction terms as possible to account for the accurate data. Our
choice \Eref{moment_scaling_fit} reflects the desire to reduce the
sensitivity of the estimate on the initial values.

It is important to note that there is a
level of ambiguity in the finite size scaling in fractal lattices,
because due to its highly irregular nature, there is, a priori, no
unique way of increasing the lattice size of a
fractal \cite{LoisonSchottePruessner:2001}. At a given level of
iterations in order to increase the lattice size further, one might
either proceed by iterating the fractal, or use the given fractal
to tessellate the hypercubic lattice of appropriate (embedding) dimension.
One might argue that finite size scaling is of course sensitive to that
choice and as a result, generates \emph{asymptotically} either the
exponents of the fractal lattice or of the embedding space. However, in
ordinary critical phenomena, there are cases
\cite{LoisonSchottePruessner:2001} where the (effective)
critical point and even the scaling functions change with the level of
iteration $i$. In the current context that translates to, for example,
the amplitude $A_q=a_q+b_q/L+c_q/L^2$ in \Eref{moment_scaling_fit} to 
acquire a dependence on $i$, which might distort the resulting estimates. 
The exponents derived above can thus 
be seen only as \emph{effective}
exponents of a fractal lattice.

\bibliography{articles,books}
\end{document}